\documentclass[10pt,twocolumn,twoside]{IEEEtran}
\pdfoutput=1
\usepackage{bbm}
\newcommand{\field}[1]{\mathbb{#1}}

\usepackage{amssymb}
\usepackage{amsmath}
\usepackage{amsfonts}
\usepackage{multirow}
\usepackage[noadjust]{cite}
\usepackage{bm}

\usepackage{dsfont}
\usepackage{yonatan}
\def\I{\ensuremath{\mathds{1}}}
\DeclareMathOperator*{\argmin}{argmin}

\ifCLASSINFOpdf
  \usepackage[pdftex]{graphicx}
   \graphicspath{{../figures/}{../jpeg/}}
  \DeclareGraphicsExtensions{.pdf,.jpeg,.png}
\else
\fi
\usepackage{array}
\usepackage{subcaption}

\usepackage{color}
\usepackage{cleveref}

\begin{document}
\title{Codebook based Audio Feature Representation for Music Information Retrieval}
\author{Yonatan~Vaizman,
        Brian~McFee,
        and~Gert~Lanckriet% <-this % stops a space
\thanks{Y. Vaizman and G. Lanckriet are with the Department
of Electrical and Computer Engineering, University of California, San Diego.}% <-this % stops a space
\thanks{B. McFee is with the Center for Jazz Studies and LabROSA, Columbia University, New York.}% <-this % stops a space
}

%\markboth{Journal of \LaTeX\ Class Files,~Vol.~6, No.~1, January~2007}%
\markboth{Transactions on Audio Speech and Language Processing}%
{Vaizman \MakeLowercase{\textit{et al.}}: Codebook based Audio Feature Representation for Music Information Retrieval}

\maketitle

\begin{abstract}
Digital music has become prolific in the web in recent decades. Automated recommendation systems are essential for users to discover music they love and for artists to reach appropriate audience. When manual annotations and user preference data is lacking (\eg\ for new artists) these systems must rely on \emph{content based} methods. Besides powerful machine learning tools for classification and retrieval, a key component for successful recommendation is the \emph{audio content representation}.

Good representations should capture informative musical patterns in the audio signal of songs. These representations should be concise, to enable efficient (low storage, easy indexing, fast search) management of huge music repositories, and should also be easy and fast to compute, to enable real-time interaction with a user supplying new songs to the system.

Before designing new audio features, we explore the usage of traditional local features, while adding a stage of encoding with a pre-computed \emph{codebook} and a stage of pooling to get compact vectorial representations. We experiment with different encoding methods, namely \emph{the LASSO}, \emph{vector quantization (VQ)} and \emph{cosine similarity (CS)}. We evaluate the representations' quality in two music information retrieval applications: query-by-tag and query-by-example. Our results show that concise representations can be used for successful performance in both applications. We recommend using top-$\tau$ VQ encoding, which consistently performs well in both applications, and requires much less computation time than the LASSO.\@
\end{abstract}

\begin{IEEEkeywords}
Music recommendation, audio representations, vector quantization, sparse coding.
\end{IEEEkeywords}

\section{Introduction}
\IEEEPARstart{I}{n} the recent decades digital music has become more accessible and music sources have become very prolific. Web servers for music exploration and recommendation contain huge repositories of music items. Hence, clever automation is required for generating good recommendation and enabling efficient search of music. Two of the most useful interfaces for a user to get music recommendations are query-by-tag (QbT) and query-by-example (QbE). In query-by-tag the system ranks music items according to relevance to a tag word (ultimately to a free-text search query), describing some semantic meaning of the desired music (emotional content, specific instruments, musical style, \etc). In query-by-example the system ranks music items according to relevance or similarity to a given music example (a song that the user already likes). This can be done in the form of an online radio, automatically creating a playlist for the user, and ultimately with an interface that enables the user to upload music clips unknown to the system and find similar music. For both these search interfaces some annotation or indexing of the songs in the repository is required. Pre-existing meta-data of the music (\eg~title, artist, lyrics, genre, instruments) is one source of such annotations and it can assist in retrieving desired items. Such information can be given with the media files as they are added to the repository (title, track duration, artist \etc), collected by experts (as done by the Music Genome Project\footnote{http://www.pandora.com/mgp.shtml}, where music experts were hired to listen to the songs and manually annotate them with relevant tags) or gathered by users of the web-service (Last.FM\footnote{http://www.last.fm/}).

Whereas the ``expert'' method to gather meta-data is labor intensive and costly, the ``user'' method is less reliable and prone to inconsistent descriptions. Another source of useful knowledge is past records of user preferences, either of specific users, for personalization purposes, or of crowds of users, for general recommendation. Such an approach is called \emph{collaborative filtering}, and it leverages co-preference of many users. For instance if many users like both artist A and artist B, and a new user likes to listen to artist A, the system will recommend artist B to that user. The collaborative filtering approach is only applicable when there is a large history of usage (plays) by many users. A recommendation system that relies solely on this approach will never suggest songs by new, unfamiliar artists, even though they are potentially suitable for some users.

Since the availability of useful meta-data and user preference data is limited, large scale music repositories must rely on \emph{content based} systems to perform efficient automatic
music recommendation. Such systems should be ``musically intelligent'', meaning they should analyze digital audio signals of music and extract meaningful information.
In the past decade much research was dedicated to constructing content based systems for music information retrieval (MIR) tasks such as music classification (to artist, genre, \etc~\cite{tzanetakis2002musical,Meng:05,reed06,ellis2007classifying,Grosse:2007,Manzagol:2008,mandel08,Joder:09,Hamel:2010,Henaff:2011,Wulfing:12,Yeh:12,yeh2013dual}), semantic annotation (auto-tagging) and retrieval (QbT~\cite{mandel06,turnbull2008semantic,eck08,mahieux08,barrington2008combining,tomasik2009,coviello2011,Nam:12,ellis2013bag}) and music similarity for song-to-song recommendation (QbE~\cite{foote1997content,logan2001music,aucouturier02,slaney08,hoffman08content,yoshii2008efficient,mcfee2012-taslp}).
The focus was mostly on machine learning algorithms that utilize basic audio features to perform the task.

In this work we use simple retrieval systems and focus on comparing different audio features and representation. Before we examine new low-level audio features, we try to make the most of traditional features, based on mel scaled spectra of short time frames. We add a stage of \emph{encoding} these frame feature vectors with a pre-computed codebook, and a stage of \emph{pooling} the coded frames (temporal integration) to get a summarized fixed-dimension representation of a whole song. The encoding detects informative local patterns and represents the frames at a higher level. The pooling stage makes the representation of a whole song compact and easy to work with (low storage, fast computation and communication), and it creates a representation that has the same dimension for all songs, regardless of their durations. We show how the same concise representation can be useful for both query-by-tag and query-by-example retrieval.

\subsection{Related work}
% Low-level local features:
Many MIR research works used mel frequency cepstral coefficients (MFCC) as audio features (\cite{foote1997content,logan2000mel,logan2001music,aucouturier02,tzanetakis2002musical,Meng:05,reed06,mandel06,ellis2007classifying,hoffman08content,eck08,mahieux08,yoshii2008efficient,Joder:09,tomasik2009,mcfee2012-taslp,Yeh:12}).
Other types of popular low-level audio features, based on short time Fourier transform are the constant-Q transform (CQT), describing a short time spectrum with logarithmically scaled frequency bins (\cite{eck08,mahieux08,Henaff:2011,Wulfing:12,Yeh:12}), and chroma features, which summarize energy from all octaves to a single 12-dimensional (per frame) representation of the chromatic scale (\cite{ellis2007classifying,barrington2008combining,bertin2012large}). While MFCC is considered as capturing timbral qualities of the sound, the CQT and chroma features are designed for harmonic properties of the music (or melodic, if using patches of multiple frames). Hamel \etal~suggested using principal component analysis (PCA) whitening of mel scaled spectral features as alternative to MFCC~\cite{hamel2011temporal}.
Some works combine heterogeneous acoustic analysis features, such as zero crossing rate, spectral flatness, estimated tempo, amplitude modulation features \etc\ (\cite{tzanetakis2002musical,mckinney2003features,flexer2006probabilistic,slaney08,Joder:09}).

% Simple temporal integration:
Low-level audio features are typically extracted from short time frames of the musical clip, then some temporal integration is done. Sometimes an early integration is performed, by taking statistics (mean, variance, covariance, \etc) of the feature vector over longer segments, or over the entire song (\eg~\cite{eck08,mandel08}). Sometimes late integration is performed, for instance: each short segment is classified and for the entire musical clip a majority vote is taken over the multiple segments' declared labels (\eg~\cite{Hamel:2010}). Such late integration systems require more computing time, since the classification operation should be done to every frame, instead of to a single vector representation per song.

% Using generative modeling:
Another approach for temporal integration is getting a compact representation of a song by generative modeling. In this approach the whole song is described
using a parametric structure that models how the song's feature vector time series was generated. Various generative models were used: GMM (\cite{aucouturier02,tzanetakis2002musical,berenzweig04,mandel06,flexer2006probabilistic,hoffman08content,turnbull2008semantic,barrington2008combining,yoshii2008efficient,Joder:09,tomasik2009})
, DTM (\cite{coviello2011}), MAR (\cite{Meng:05,Joder:09}), ARM (\cite{coviello_vaizman2012}), HMM (\cite{reed06,Joder:09,coviello2012}), HDP (\cite{hoffman08content}). Although these models have been shown very useful and some of them are also time-efficient to work with, the representation of a song using a statistical model is less convenient than a vectorial representation. The former requires retrieval systems that fit specifically to the generative model while the later can be processed by many generic machine learning tools. Computing similarity between two songs is not straight forward using a generative model (although there are some ways to handle it, like the probability product kernel (\cite{Jebara:04,barrington2008combining,coviello_vaizman2012})), whereas for vectorial representation there are many efficient generic ways to compute similarity between two vectors of the same dimension. In~\cite{coviello_vaizman2012} the song level generative model (multivariate autoregressive mixture) was actually used to create a kind of vectorial representation for a song by describing the frequency response of the generative model's dynamic system, but still, being a mixture model, the resulted representation was a \emph{bag} of four vectors, and not a single vectorial representation.

% Encoding low-level features with codebook:
Encoding of low-level features using a pre-calculated codebook was examined for audio and music. Quantization tree (\cite{foote1997content}), vector quantization (VQ) (\cite{reed06,Lyon:2010,mcfee2012-taslp}), sparse coding with the LASSO (\cite{Grosse:2007}) and other variations (\cite{Wulfing:12,Henaff:2011}) were used to represent the features at a higher level. Sparse representations were also applied directly to time domain audio signals, with either predetermined kernel functions (Gammatone) or with a trained codebook (\cite{Smith:2006,Manzagol:2008}). As alternative to the heavy computational cost of solving optimization criteria (like the LASSO) greedy algorithms like matching pursuit have also been applied (\cite{Smith:2006,Manzagol:2008,Lyon:2010}).

% Mixing of models and deep coding:
Heterogeneous and multi-layer systems have been proposed. The bag of systems approach combined various generative models as codewords (\cite{ellis2013bag}). Multi-modal signals (audio and image) were combined in a single framework (\cite{Yang:12}). Even the codebook training scheme, which was usually unsupervised, was combined with supervision to get a boosted representation for a specific application (\cite{Yang:12,Yeh:12}).
Deep belief networks were used in~\cite{Hamel:2010}, also combining unsupervised network weights training with supervised fine tuning. In~\cite{yeh2013dual} audio features were processed in two layers of encoding with codebooks.

% Comparing encoding methods:
Several works invested in comparing different encoding schemes for audio, music and image.
Nam \etal~examined different variations of low-level audio processing, and compared different encoding methods (VQ, the LASSO and sparse restricted Boltzman machine) for music annotation and retrieval with the CAL500 dataset~\cite{Nam:12}.
Yeh \etal~reported to find superiority of sparsity-enforced dictionary learning and $L1$-regularized encoding over regular VQ for genre classification.
In~\cite{Coates:2011} Coates and Ng examined the usage of different combinations of dictionary training algorithms and encoding algorithms to better explain the successful performance of sparse coding in previous works. They concluded that the dictionary training stage has less of an impact on the final performance than the encoding stage and that the main merit of sparse coding may be due to its nonlinearity, which can be achieved also with simpler encoders that apply some nonlinear soft thresholding. In~\cite{coates2010analysis} Coates \etal~examined various parameters of early feature extraction for images (such as the density of the extracted patches) and showed that when properly extracting features, one can use simple and efficient algorithms (k-means clustering and single layer neural network) and achieve image classification performance as high as other, more complex systems.

\subsection{Our contribution}

In this work we look for \emph{compact} audio content representations that will be powerful for two different MIR applications: query-by-tag and query-by-example. We perform a \emph{large scale} evaluation, using the CAL10k and Last.FM datasets.
We assess the effect of various design choices in the ``low-level-feature, encoding, pooling'' scheme, and eventually recommend a representation ``recipe'' (based on vector quantization) that is efficient to compute, and has consistent high performance in both MIR applications.

The remainder of the paper is organized as follows: in \Cref{sec:song_representation} we describe the audio representations that we compare, including the low-level audio features, the encoding methods and pooling. In \Cref{sec:mir_tasks} we describe the MIR tasks that we evaluate, query-by-tag and query-by-example retrieval. In \Cref{sec:experimental_setup} we specify the dataset used, the data processing stages and the experiments performed. In \Cref{sec:results} we describe our results, followed by conclusions in \Cref{sec:conclusions}.

\section{Song representation}\label{sec:song_representation}
We examine the encoding-pooling scheme to get a compact representation for each song (or musical piece). The scheme is comprised of three stages:
\begin{enumerate}
\item \textbf{Short time frame features:} each song is processed to a time series of low-level feature vectors, $X \in \field{R}^{d \times T}$ ($T$ time frames, from each a $d$ dimensional feature vector is extracted).
\item \textbf{Encoding:} each feature vector $x_t \in \field{R}^{d}$ is then encoded to a code vector $c_t \in \field{R}^{k}$, using a pre-calculated dictionary $D \in \field{R}^{d \times k}$, a codebook of $k$ ``basis vectors'' of dimension $d$. We get the encoded song $C \in \field{R}^{k \times T}$.
\item \textbf{Pooling:} the coded frame vectors are pooled together to a single compact vector $v \in \field{R}^{k}$.
\end{enumerate}
This approach is also known as the bag of features (BoF) approach: where features are collected from different patches of an object (small two-dimensional patches of an image, short time frames of a song, \etc) to form a variable-size set of detected features.
The pooling stage enables us to have a unified dimension to the representations of all songs, regardless of the songs' durations. A common way to pool the low-level frame vectors together is to take some statistic of them, typically their mean. For a monotonic, short song, such a statistic may be a good representative of the properties of the song.

However, a typical song is prone to changes in the spectral content, and a simple statistic pooling function over the low-level feature frames may not represent it well. For that reason the second stage (encoding) was introduced. In a coded vector, each entry encodes the presence/absence/prominence of a specific pattern in that frame. The pre-trained codebook holds codewords (patterns) that are supposed to roughly represent the variety of prominent patterns in songs. The use of sparsity in the encoding (having only \emph{few} basis vectors active in each frame), promotes selecting codewords that represent typical whole sound patterns (comprised of possibly many frequency bands). The pooling of these coded vectors is meaningful: using mean pooling gives a histogram representation, stating the frequency of occurrence of each sound pattern, while using max-abs (maximum absolute value) pooling gives more of an indication representation --- for each sound pattern, did it appear anytime in the song, and in what strength. For some encoding methods it is appropriate to take absolute value and treat negative values far from zero as strong values. In our experiments we used three encoding systems, the LASSO (\cite{Tibshirani:96}), vector quantization (VQ), and cosine similarity (CS) (all explained later), and applied both mean and max-abs pooling functions to the coded vectors.

\subsection{Low-level audio features}
In this work we use spectral features that are commonly assumed to capture timbral qualities. Since we are not interested in melodic or harmonic information, but rather general sound similarity, or semantic representation, we assume timbral features to be appropriate here (an assumption that is worth examination). Our low-level features are based on mel frequency spectra (MFS), which are calculated by computing the short time Fourier transform (STFT), summarizing the spread of energy along mel scaled frequency bins, and compressing the values with logarithm. Mel frequency cepstral coefficients (MFCCs~\cite{logan2000mel}) are the result of further processing MFS, using discrete cosine transform (DCT), in order to both create uncorrelated features from the correlated frequency bins, and reduce the feature dimension. In addition to the traditional DCT we alternatively process the MFS with another method for decorrelating, based on principal component analysis (PCA). Processing details are specified in \Cref{subsec:processing}.

\subsection{Encoding with the LASSO}
The least absolute shrinkage and selection operator (the LASSO) was suggested as an optimization criterion for linear regression that selects only few of the regression coefficients to have effective magnitude, while the rest of the coefficients are either shrunk or even nullified~\cite{Tibshirani:96}. The LASSO does that by balancing between the regression error (squared error) and an $L1$ norm penalty over the regression coefficients, which typically generates sparse coefficients. Usage of the LASSO's regression coefficients as a representation of the input is often referred to as ``sparse coding''.
In our formulation, the encoding of a feature vector $x_t$ using the LASSO criterion is:

\begin{align*}
& c_t = \argmin_{c \in \field{R}^{k}}{\frac{1}{2}\parallel x_t - D c\parallel_2^2 + \lambda \parallel c\parallel_1}. \\
\end{align*}

Intuitively it seems that such a sparse linear combination might represent separation of the music signal to meaningful components (\eg~separate
instruments). However, this is not necessarily the case since the LASSO allows coefficients to be negative and the subtraction of
codewords from the linear combination has little physical interpretability when describing how musical sounds are generated.
To solve the LASSO optimization problem we used the alternating direction method of multipliers (ADMM) algorithm. The general algorithm, and a specific version for the LASSO are detailed in~\cite{Boyd:10}.
The $\lambda$ parameter can be interpreted as a sparsity parameter: the larger it is, the more weight will be dedicated to the $L1$ penalty, and the resulted code will typically be more sparse.

\subsection{Encoding with vector quantization (VQ)}
In vector quantization (VQ) a continuous multi-dimensional vector space is quantized to a discrete finite set of bins, each having its own representative vector. The training of a VQ codebook is essentially a clustering that describes the distribution of vectors in the space. During encoding, each frame's feature vector $x_t$ is quantized to the closest codeword in the codebook, meaning it is encoded as $c_t$, a sparse binary vector with just a single ``on'' value, in the index of the codeword that has smallest distance to it (we use Euclidean distance).
It is also possible to use a softer version of VQ, selecting for each feature vector $x_t$ the $\tau$ nearest neighbors among the $k$ codewords, creating a code vector $c_t$ with $\tau$ ``on'' values and $k-\tau$ ``off'' values:

\begin{align*}
& c_t(j) = \frac{1}{\tau}\I \left[D_j \in \tau\text{-nearest neighbors of } x_t\right], \\
& j \in \{1,2,\ldots ,k\}.
\end{align*}

Such a soft version can be more stable: whenever a feature vector has multiple codewords in similar vicinity (quantization ambiguity), the hard threshold of selecting just one codeword will result in distorted, noise-sensitive code, while using top-$\tau$ quantization will be more robust. This version also adds flexibility and richness to the representation: instead of having $k$ possible codes for every frame, we get $\binom{k}{\tau}$ possible codes. Of course, if $\tau$ is too large, we may end up with codes that are trivial --- all the songs will have similar representations and all the distinguishing information will be lost. The sparsity parameter $\tau$ here is actually a density parameter, with larger values causing denser codes. By adjusting $\tau$ we can directly control the level of sparsity of the code, unlike in the LASSO, where the effect of adjusting the parameter $\lambda$ is indirect, and depends on the data. The values in the coded vectors are binary (either $0$ or $\frac{1}{\tau}$). Using max-abs pooling on these code vectors will result in binary final representations. Using mean pooling results in a codeword histogram representation with richer values. We only use mean pooling for VQ in our experiments.

In~\cite{mcfee2012-taslp} it was shown that for codeword histogram representations (VQ encoding and mean pooling), it was beneficial to take the square root of every entry, consequently transforming the song representation vectors from points on a simplex ($\sum\limits_{j=1}^k |v_j|=1$) to points on the positive orthant of a sphere ($\sum\limits_{j=1}^k |v_j|^2=1$). The authors called it PPK transformation, since a dot product between two transformed vectors is equivalent to the probability product kernel (PPK) with power 0.5 on the original codeword histograms~\cite{Jebara:04}. We also experiment with the PPK-transformed versions of the codeword histogram representations.

\subsection{Encoding with cosine similarity (CS)}
VQ encoding is simple and fast to compute (unlike the LASSO, whose solving algorithms, like ADMM, are iterative and slow). However, it involves a hard threshold (even when $\tau>1$) that possibly distorts the data and misses important information. When VQ is used for communication and reconstruction of signal it is necessary to use this thresholding in order to have a low bit rate (transmitting just the index of the closest codeword).

However, in our case of encoding songs for retrieval we have other requirements. As an alternative to VQ we experiment with another form of encoding, where
each dictionary codeword is being used as a linear filter over the feature vectors: instead of calculating the \emph{distance} between each feature vector and each codeword (as done in VQ), we calculate a \emph{similarity} between them --- the (normalized) dot product between the feature vector and the codeword: $\frac{\langle x_t,D_j\rangle}{\|x_t\|_2}$. Since the codewords we trained are forced to have unit $L2$ norm, this is equivalent to the cosine similarity (CS). The codewords act as pattern matching filters, where frames with close patterns get higher response.
%% Janani suggested perhaps to put the formula in a separate line instead of embedded in text line.

For the CS encoding we use the same codebooks that are used for VQ.\@ For each frame, selecting the closest (by Euclidean distance) codeword is equivalent to selecting the codeword with largest CS with the frame. So CS can serve as a softer version of VQ.\@ The $L2$ normalization of each frame (to get CS instead of unnormalized dot product) is important to avoid having a bias towards frames that have large magnitudes, and can dominate over all other frames in the pooling stage. In our preliminary experiments we verified that this normalization is indeed significantly beneficial to the performance. The CS regards only to the ``shape'' of the pattern but not to its magnitude and gives a fair ``vote'' also to frames with low power. Unlike the unnormalized dot product the response values of CS are limited to the range $[-1,1]$, and are easier to interpret and to further process.

In the last stage of the encoding we introduce non-linearity in the form of the shrinkage function $y(x)=\text{sign}(x)*\max(|x|-\theta,0)$ (values with magnitude less than $\theta$ are nullified and larger magnitude values remain with linear, but shrinked, response). Using $\theta=0$ maintains the linear responses of the filters, while $\theta>0$ introduces sparsity, leaving only the stronger responses. Such a nonlinear function is sometimes called ``soft thresholding'' and was used in various works before to compete with the successful ``sparse coding'' (the LASSO) in a fast feed-forward way (\cite{Coates:2011}).

\subsection{Dictionary training}
The training of the dictionaries (codebooks) is performed with the online learning algorithm for sparse coding presented by Mairal \etal\ (\cite{Mairal:10}). As an initialization stage we apply online k-means to a stream of training $d$-dimensional feature vectors, to cluster them to an initial codebook of $k$ codewords. This initial dictionary is then given to the online algorithm, which alternates between encoding a small batch of new instances using the current dictionary, and updating the dictionary using the newly encoded instances. In each iteration the updated codewords are normalized to have unit $L2$ norm.

\section{MIR tasks}\label{sec:mir_tasks}
We examine the usage of the various song representations for two basic MIR applications, with the hope to find stable representations that are consistently successful in both tasks. We use simple, linear machine learning methods, seeing as our goal here is finding useful song representations, rather than finding sophisticated new learning algorithms.

\subsection{Query-by-tag (QbT)}
We use $L2$-regularized logistic regression as a tag model. For each semantic tag we use the positively and negatively labeled training instances ($k$-dimensional song vectors) to train a tag model. Then for each song in the test set and for each tag we use the trained tag model to estimate the probability of the tag being relevant to the song (the likelihood of the song-vector given the tag model).
For each song, the vector of tag-likelihoods is then normalized to be a categorical probability over the tags, also known as the semantic multinomial (SMN) representation of a song.

Retrieval: For each tag the songs in the test set are ranked according to their SMN value relevant to the tag. Per-tag area under curve (AUC), precision at top-10 (P@10) and average precision (AP) are calculated as done in~\cite{turnbull2008semantic,coviello2011}.
These per-tag scores are averages over the tags to get a general score (mean (over tags) AP is abbreviated MAP).

\subsection{Query-by-example (QbE)}
Given a query song, whose audio content is represented as vector $q \in \field{R}^k$, our query-by-example system calculates its distance $dist(q,r)$ from each repository song $r \in \field{R}^k$ and the recommendation retrieval result is the repository songs ranked in increasing order of distance from the query song. The Euclidean distance is a possible simple distance measure between songs' representations. However, it grants equal weight to each of the vectors' dimensions, and it is possible that there are dimensions that carry most of the relevant information, while other dimensions carry just noise. For that reason, we use a more general metric as distance measure, the Mahalanobis distance: $dist(q,r)=\sqrt{{(q-r)}^T W (q-r)}$, when $W \in \field{R}^{k\times k}$ is the parameter matrix for the metric ($W$ has to be a positive semidefinite matrix for a valid metric).

In~\cite{mcfee10_mlr} McFee \etal~presented a framework for using a metric for query-by-example recommendation systems, and a learning algorithm --- metric learning to rank (MLR) --- for training the metric parameter matrix $W$ to optimize various ranking performance measures.
In~\cite{mcfee2012-taslp} the authors further demonstrated the usage of MLR for music recommendation, and the usage of collaborative filtering data to train the metric, and to test the ranking quality. Here we followed the same scheme: collaborative filtering data are used to define artist-artist similarity (or relevance), and song-song binary relevance labels. MLR is then applied to training data to learn a metric $W$. The learnt metric is tested on a test set. Further details are provided in \Cref{subsec:processing}. Same as for query-by-tag, we apply the same scheme to different audio content representations and compare the performance of query-by-example.

\section{Experimental setup}\label{sec:experimental_setup}
\subsection{Data}\label{subsec:data}
In this work we use the CAL10k dataset~\cite{Tingle10}. This dataset contains $10,865$ full-length songs from over $4,500$ different artists, ranging over $18$ musical genres. Throughout the paper we use the convenient term ``song'' to refer to a music item/piece (even though many of the items in CAL10k are pieces of classical music and would commonly not be called songs). It also contains semantic tags harvested from the Pandora website, including $475$ acoustic tags and $153$ genre (and sub-genre) tags. These tag annotations were done by humans, musical experts. The songs in CAL10k are weakly labeled in the sense that if a song doesn't have a certain tag, it doesn't necessarily mean that the tag is not relevant for the song, but for evaluation we assume missing song-tag associations to be negative labels. We filter the tags to include only the $581$ tags that have at least $30$ songs associated with them.

For the query-by-example task we work with the intersection of artists from CAL10k and the Last.FM collaborative filter data, collected by Celma (\cite{celma:2010} chapter 3). As done in~\cite{mcfee2012-taslp} we calculate the artist-artist similarity based on Jaccard index (\cite{jaccard}) and the binary song-song relevance metric, which is used as the target metric to be emulated by MLR.\@

For the dictionary training we use external data --- $\sim3500$ audio files of songs/clips by $\sim700$ artists that \emph{do not appear} in CAL10k. These clips were harvested from various interfaces on the web and include both popular and classical music. This is unlike the sampling from \emph{within} the experimental set, as was done in~\cite{Nam:12}, which might cause over-fitting.

\subsection{Processing}\label{subsec:processing}
Audio files are averaged to single channel (in case they are given in stereo) and re-sampled at $22,050$Hz. Feature extraction is done over half-overlapping short frames of $2,048$ samples (a feature vector once every $1,024$ samples, which is once every $\sim 46msec$). The magnitude spectrum (magnitude DFT) of each frame is summarized into $34$ Mel-scaled frequency bins, and log value is saved to produce initial MFS features. To get the MFCC features a further step of discrete cosine transform (DCT) is done and $13$ coefficients are saved. The $1^{st}$ and $2^{nd}$ instantaneous derivatives are augmented to produce MFCC$\Delta$ ($d=39$) and MFS$\Delta$ ($d=102$) feature vectors. The next step is to standardize the features so that each dimension would have zero mean and unit variance (according to estimated statistics). In order to have comparable audio features, we reduce the dimension of the MFS$\Delta$ to $39$ dimensions using a PCA projection matrix (pre-estimated from the dictionary training data) to get MFS$\Delta$PC features.

The dictionary training set is used to both estimate statistics over the raw features, and to train the dictionary: first the mean vector and vector of standard deviation of each dimension are calculated over the pool of low-level feature vectors (either MFCC$\Delta$ or MFS$\Delta$). Then all the vectors are standardized (by subtracting the mean vector and dividing each dimension by the appropriate standard deviation) to get the pool of standardized feature vectors. For the MFS$\Delta$ another stage of PCA projection is done (using a projection matrix that was also estimated from the same training set). From each training audio file a segment of $20\sec$ is randomly selected, processed and its feature vectors are added to a pool of vectors (resulting in $1.5~\text{million}$ vectors), which are scrambled to a random order and fed to the online dictionary training algorithm.

For each codebook size $k$ the LASSO codebook is trained with $\lambda=1$ (this codebook is later used for the LASSO encoding with various values of $\lambda$) and the VQ codebook is trained with $\tau=1$ (this codebook is later used for VQ encoding with various valued of $\tau$ and for CS encoding).
%\bn{Is this fair? Why $\lambda=1$?} \bn{Is this fair? Why $\lambda=1$?}

For training the logistic regression model of a tag, an internal cross validation is done over different combinations of parameters (weight of regularization, weight of false negative error, weight of false positive error), each of which could take values of $\left[0.1,1,10,100\right]$.
This cross validation is done using only the training set, and the parameter set selected is the one that optimizes the AUC.\@ After selecting the best parameter set for a tag, the entire training set is used to train the tag model with these parameters.

The query-by-tag evaluation is done with 5-fold cross validation. For each fold no artist appears in both the train set and the test set. The performance scores that were averaged over tags in each fold, are then averaged over the five folds.
The query-by-example evaluation is done with 10 splits of the data in the same manner as done in~\cite{mcfee2012-taslp}. We use the AUC rank measure to define the MLR loss between two rankings (marked as $\Delta(y*,y)$ in~\cite{mcfee10_mlr}). For each split we train $W$ over the train set with multiple values of the slack trade off parameter $C$ ($10^{-2},10^{-1},\ldots,10^8$) and for each value test the trained metric on the validation set. The metric that results in highest AUC measure on the validation set is then chosen and tested on the test set. We report the AUC results on the test set, averaged over the 10 splits.

For QbE PCA decorrelation and dimensionality reduction is performed on the data: in each split the PCA matrix is estimated from the train set and the song representation vectors (of train, validation and test set) are projected to a predetermined lower dimension (so the trained matrices $W$ are in fact not $(k\times k)$ but smaller). In~\cite{mcfee2012-taslp} the heuristic was to reduce to the estimated effective dimensionality --- meaning to project to the first PCs covering 0.95 of the covariance (as estimated from the train set). However, in our experiments we noticed that reducing to the effective dimensionality caused deterioration of performance when the effective dimensionality decreased, while keeping a fixed reduction-dimension kept stable or improving performance. So keeping 0.95 of the covariance is not the best practice. Instead, for every $k$ we fix the dimension to reduce to (across different encoders and encoding parameters).

When testing each of the 10 splits, each song in the query set (either the validation set or the test set) is used as a query to retrieve relevant songs from the train set --- the train songs are ranked according to the trained metric and the ranking for the query song is evaluated (AUC score). The average over query songs is then taken.

\subsection{Experiments}
Each experiment regards to a different type of audio-content representation. We experiment with different combinations of the following parameters:
\begin{itemize}
  \item low-level features: MFCC$\Delta$ or MFS$\Delta$PC,\@
  \item codebook size $k \in \{128,256,512,1024\}$,
  \item encoding method: the LASSO, VQ or CS,\@
  \item encoding parameters:
    \begin{itemize}
      \item the LASSO:\ $\lambda \in \{0.01,0.1,0.5,1,2,10,100\}$,
      \item VQ:\ $\tau \in \{1,2,4,8,16,32\}$,
      \item \emph{CS}:\ $\theta \in \{0,0.1,0.2\ldots,0.9\}$,
    \end{itemize}
  \item pooling function: either mean or max-abs,
  \item VQ:\ either using PPK-transformation or not.
\end{itemize}

\section{Results}\label{sec:results}
\subsection{query-by-tag results}
First, for comparison, we present baseline results: chance level scores are the result of scrambling the order of song labels and performing the query-by-tag task, while using the representations with MFS$\Delta$PC, $k=1024$ and VQ encoding with $\tau=8$. Then, to control for the encoding methods in our scheme, we perform the experiments without the encoding stage (instead of encoding the feature vectors with a codebook, leaving them as low-level features and pooling them) for both the MFCC$\Delta$ and MFS$\Delta$PC low-level features. Finally, as alternative to the codebook based systems, we evaluate the HEM-GMM system, which is the suitable candidate from the generative models framework, being computation efficient and assuming bag of features (like our current codebook systems). We process the data as was done in~\cite{coviello2011} for HEM-GMM, using our current 5-fold partition. \Cref{tab:baseline_results} presents these baselines.

\begin{table}[h] %\footnotesize
\centering
\begin{tabular}{ | r|c | c | c c c|}
				\cline{4-6}
                                \multicolumn{3}{c|}{}   & P@10 & MAP & AUC \\  \hline
   \multicolumn{3}{|r|}{chance level}                  & 0.02 & 0.02 & 0.5\\ \hline
   \multirow{5}{*}{no encoding}     & audio feature & pooling &  &  &  \\  \cline{2-6}
                                & MFCC$\Delta$ & mean  & 0.09 & 0.07 & 0.76\\
                                & MFCC$\Delta$ & max-abs & 0.09 & 0.07 & 0.75 \\ \cline{2-6}
                                & MFS$\Delta$PC & mean & 0.10 & 0.08 & 0.77 \\
                                & MFS$\Delta$PC & max-abs & 0.09 & 0.07 & 0.75 \\
				\hline
                                & \multicolumn{2}{c|}{HEM-GMM} & 0.21 & 0.16 & 0.84 \\
				\hline
\end{tabular}
  \caption{Query-by-tag --- baseline results}
\label{tab:baseline_results}
\end{table}
%\vspace{12pt}

\begin{figure}
    \centering
    \includegraphics[width=0.45\textwidth]{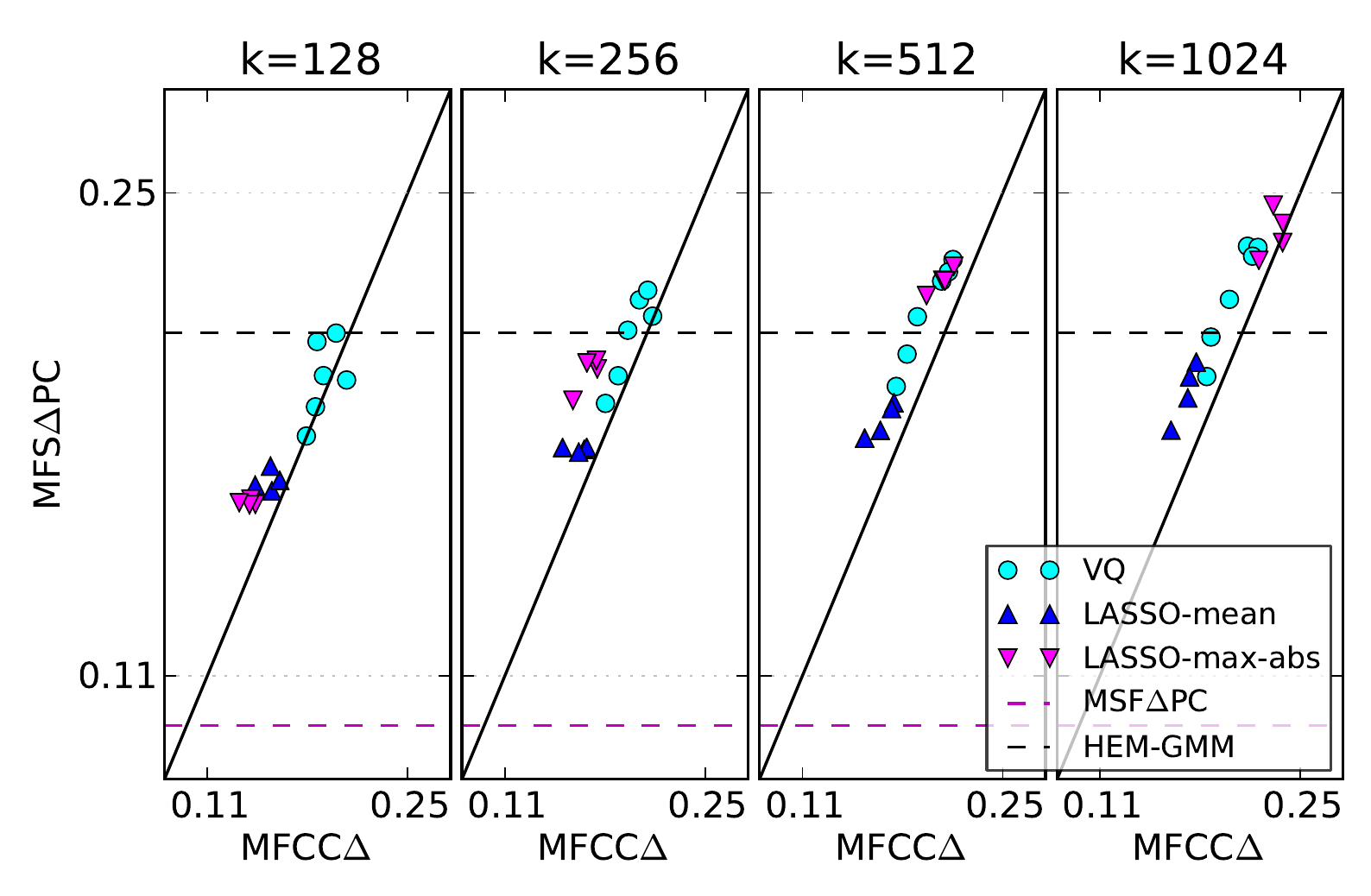}
    \caption{Comparison of the two low-level audio features. Each point regards to a specific combination of encoder, encoding parameter and pooling, and displays the performance score (QbT P@10) when using MFCC$\Delta$ (x-axis) and MFS$\Delta$PC (y-axis) as low-level features.}
\label{fig:mfcc_vs_mfs}
\end{figure}

\begin{figure}
    \centering
    \begin{subfigure}[b]{0.5\textwidth}
        \centering
        \includegraphics[width=\textwidth]{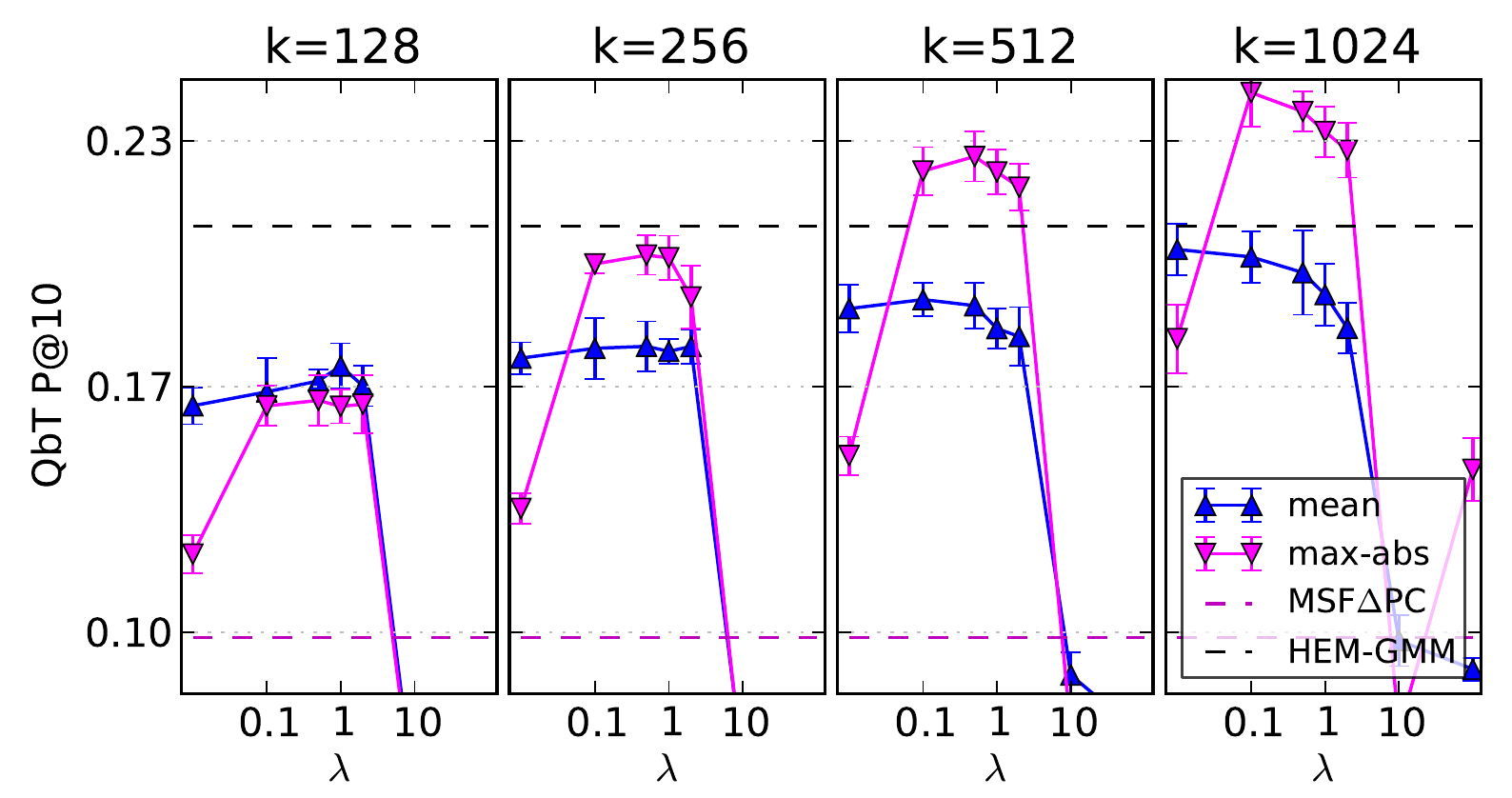}
        \caption{Query-by-tag with the LASSO.}
\label{fig:sparsity_admm}
    \end{subfigure}
    \begin{subfigure}[b]{0.5\textwidth}
        \centering
        \includegraphics[width=\textwidth]{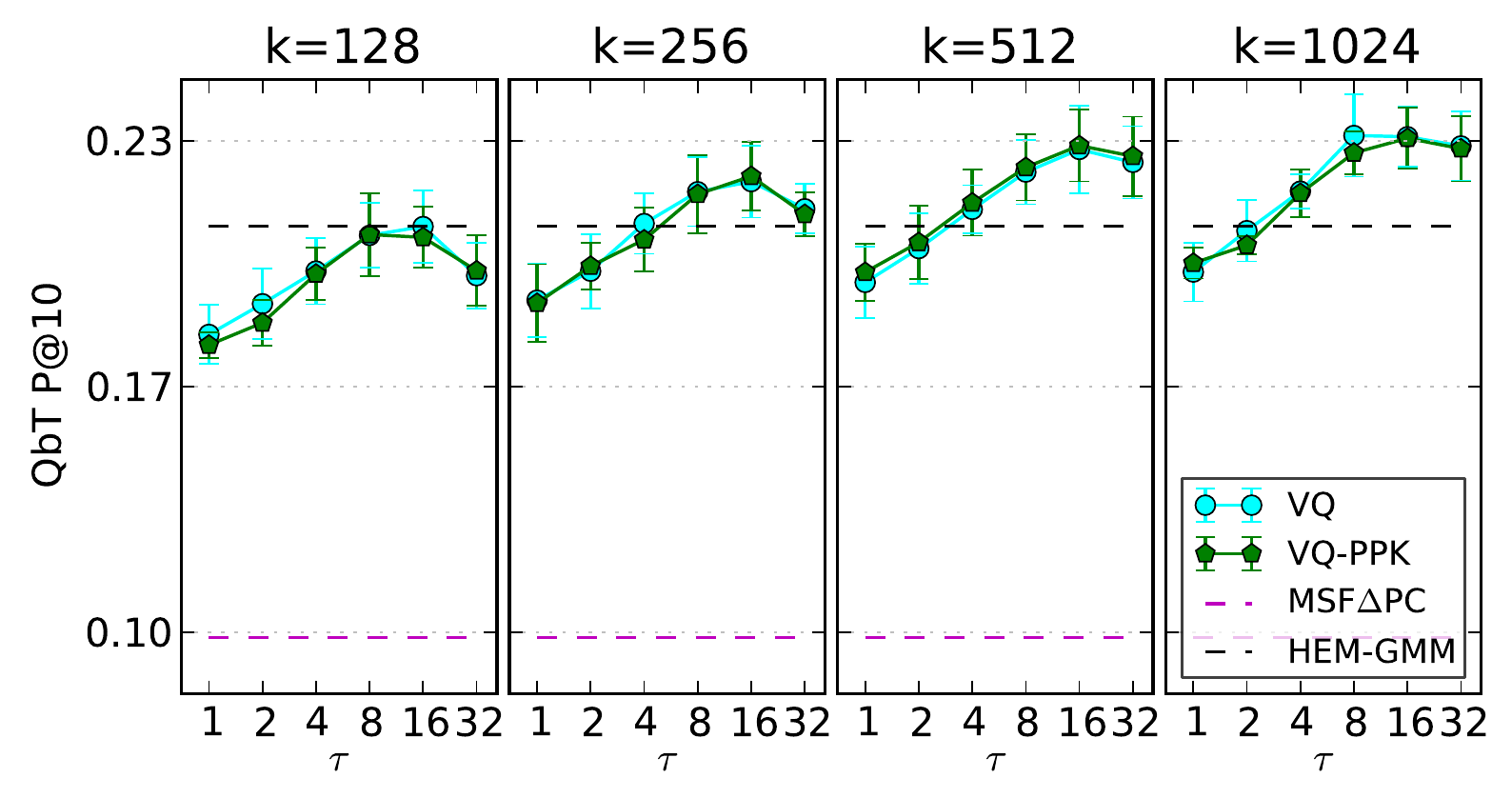}
        \caption{Query-by-tag with VQ.}
\label{fig:sparsity_vq}
    \end{subfigure}
    \begin{subfigure}[b]{0.5\textwidth}
        \centering
        \includegraphics[width=\textwidth]{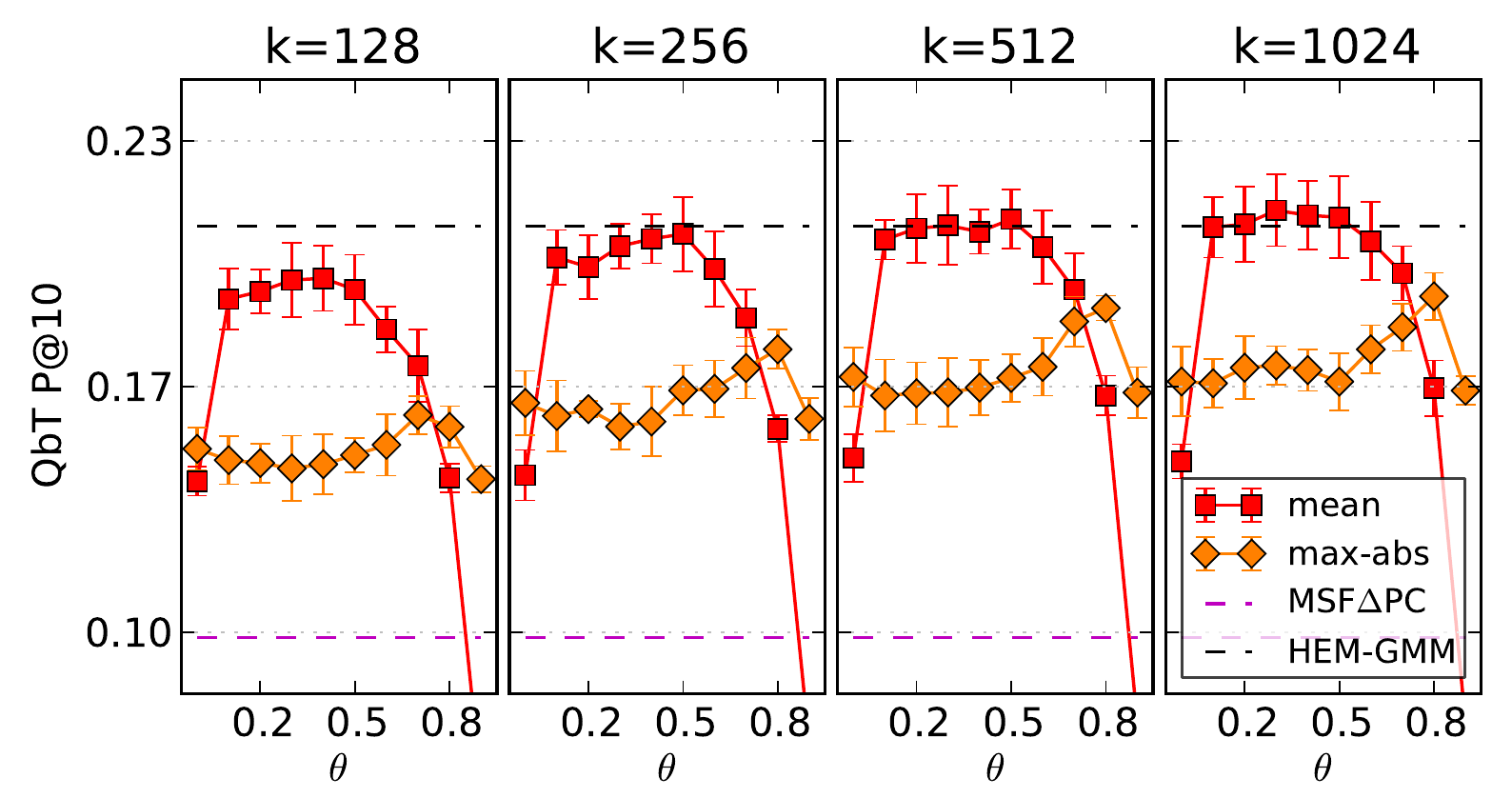}
        \caption{Query-by-tag with CS.}
\label{fig:sparsity_linear_filters}
    \end{subfigure}

\caption{Query-by-tag with different encoders. Effect of pooling or PPK-transformation (shape) and sparsity parameter (x-axis): $\lambda$ (log-scale) for the LASSO (a), $\tau$ (log-scale) for VQ (b) and $\theta$ for CS (c). Error bars indicate one standard deviation over the five folds.}
\label{fig:qbt}
\end{figure}

In \Cref{fig:mfcc_vs_mfs,fig:qbt} we show plots for the P@10 rank measure (this measure is the more practical objective, since in real recommendation systems, the user typically only looks at the top of the ranked results). Graphical results for the other performance measures are provided in the supplementary material.
In some plots error bars are added: the error bars represent the standard deviation of the score (over the five folds for query-by-tag, and over the 10 splits for query-by-example).

\textbf{Low-level features:} \Cref{fig:mfcc_vs_mfs} shows the query-by-tag results for comparison between the two low-level features: MFCC$\Delta$ and MFS$\Delta$PC.\@ Each point in the graphs compares the performance using MFCC$\Delta$ (x-axis) to the performance using MFS$\Delta$PC (y-axis), when all the other parameters ($k$, encoding method, encoding parameter, pooling method) are the same. Multiple points with the same shape represent experiments with the same encoder and pooling, but different encoding parameter. The main diagonal line ($y=x$) is added to emphasize the fact that in the majority of the experiments performance with MFS$\Delta$PC was better than MFCC$\Delta$.\@ Statistical tests (paired two-tailed t-test between two arrays of $\sim2900$ per-fold-per-tag scores) support the advantage of MFS$\Delta$PC:\ most comparisons show statistically significant advantage to MFS$\Delta$PC (all except six points on the plots. P-value well below 0.05), and only one point (for $k=128$ with VQ and $\tau=32$) has significant advantage to MFCC$\Delta$.

While it is expected that the data-driven decorrelation (PCA) performs better than the predetermined projection (DCT), it is interesting to see that the difference is not so dramatic (points are close to the main diagonal) --- MFCC managed to achieve performance close to the data-trained method. Other than the advantage of training on music data, another explanation to the higher performance of MFS$\Delta$PC can be the effect of first taking a local dynamic structure (concatenating the ``deltas'' to the features) and only then decorrelating the features-$\Delta$ version (as we did here for MFS$\Delta$PC).

These results also demonstrate the advantage of using \emph{some} encoding over low-level features before pooling them: all these performances (for both MFCC$\Delta$ and MFS$\Delta$PC) are better than the baseline results with no encoding (\Cref{tab:baseline_results}. The highest of the ``no encoding'' baselines is also added as reference line in the plots). We can also notice the improvement with increasing codebook sizes (the different subplots). Similar results are seen for the other performance measures (AUC, MAP) --- graphs shown in the supplementary material. The remainder of the results focus on the MFS$\Delta$PC low-level features.

\textbf{The LASSO encoding:} \Cref{fig:sparsity_admm} shows the query-by-tag results (P@10) with MFS$\Delta$PC features for the LASSO encoding.
The LASSO is sensitive to the value of its sparsity parameter $\lambda$. When $\lambda$ is too high (in this case $\lambda=10,100$), the resulted code is too sparse and loses important information, causing deteriorated performance. When $\lambda$ is too small ($0.01$) the code is too dense. This doesn't effect much when using mean pooling, but harms performance for the max-abs pooling representation. Similar results are seen for AUC and MAP measures (supplementary material). There seems to be an advantage to using max-abs pooling over mean pooling, however this advantage is not apparent for the smaller codebook size (128) and not in the AUC performance.

\textbf{VQ encoding:} \Cref{fig:sparsity_vq} shows the results (P@10) with MFS$\Delta$PC features and for VQ encoding. These results depict a clear effect of the VQ density parameter $\tau$: ``softening'' the VQ by quantizing each frame to more than one codeword significantly improves the performance. There is an optimal peak for $\tau$, typically at 8 or 16 --- increasing $\tau$ further causes performance to deteriorate, especially with a small codebook.
The effect of the PPK-transformation is small and inconsistent. These trends are consistent also for AUC and MAP (supplementary material).

\textbf{Cosine similarity encoding:}
The query-by-tag results (P@10) for CS encoding (\Cref{fig:sparsity_linear_filters}) demonstrate the effect of adjusting the sparsity parameter $\theta$ (the ``knee'' of the shrinkage function): the optimal value is not too small and not too large. This is more dramatically seen for the mean pooling: there is a significant advantage in adding \emph{some} non-linearity (having $\theta>0$), and at the other end having the code too sparse ($\theta$ too large) causes a drastic reduction in performance. For max-abs pooling, generally performance was not as good as mean pooling, having a sharp peak at $\theta=0.8$.

\begin{table*} \footnotesize
\centering
\begin{tabular}{ |l|l|l|l| l l l|}
\hline
	 \multicolumn{4}{|c|}{representation} & \multicolumn{3}{c|}{QbT}\\
	 k & encoding & parameter & pooling & P@10 & MAP & AUC\\ \hline
	1024 & VQ (with PPK) & $\tau=8$ & mean  & $0.230$ \tiny{$(9\text{e}-08)$} & $0.185$ \tiny{$(1\text{e}-09)$} & $0.867$ \tiny{$(9\text{e}-06)$} \\
	1024 & VQ (no PPK) & $\tau=8$ & mean  & $0.235$ \tiny{$(1\text{e}-04)$} & $0.188$ \tiny{$(4\text{e}-06)$} & $0.868$ \tiny{$(2\text{e}-04)$} \\
	1024 & the LASSO & $\lambda=0.1$ & max-abs  & \textbf{0.246} & \textbf{0.195} & \textbf{0.874} \\
	1024 & cosine similarity & $\theta=0.4$ & mean  & $0.212$ \tiny{$(8\text{e}-27)$} & $0.175$ \tiny{$(1\text{e}-34)$} & $0.863$ \tiny{$(2\text{e}-19)$} \\
	1024 & cosine similarity & $\theta=0.8$ & max-abs  & $0.190$ \tiny{$(9\text{e}-62)$} & $0.156$ \tiny{$(2\text{e}-106)$} & $0.852$ \tiny{$(2\text{e}-82)$} \\
	512 & VQ (with PPK) & $\tau=8$ & mean  & $0.226$ \tiny{$(6\text{e}-11)$} & $0.181$ \tiny{$(2\text{e}-17)$} & $0.867$ \tiny{$(3\text{e}-07)$} \\
	512 & the LASSO & $\lambda=0.1$ & max-abs  & $0.225$ \tiny{$(2\text{e}-13)$} & $0.176$ \tiny{$(6\text{e}-39)$} & $0.862$ \tiny{$(3\text{e}-46)$} \\
	256 & VQ (with PPK) & $\tau=8$ & mean  & $0.218$ \tiny{$(4\text{e}-19)$} & $0.176$ \tiny{$(5\text{e}-31)$} & $0.863$ \tiny{$(2\text{e}-14)$} \\
	256 & the LASSO & $\lambda=0.1$ & max-abs  & $0.199$ \tiny{$(1\text{e}-53)$} & $0.153$ \tiny{$(2\text{e}-129)$} & $0.840$ \tiny{$(7\text{e}-230)$} \\
	128 & VQ (with PPK) & $\tau=8$ & mean  & $0.207$ \tiny{$(8\text{e}-36)$} & $0.165$ \tiny{$(3\text{e}-65)$} & $0.857$ \tiny{$(2\text{e}-32)$} \\
	128 & the LASSO & $\lambda=0.1$ & max-abs  & $0.160$ \tiny{$(6\text{e}-142)$} & $0.122$ \tiny{$(9\text{e}-261)$} & $0.811$ \tiny{$(0\text{e}+00)$} \\
\hline
 	 \multicolumn{4}{|c|}{HEM-GMM} & $0.210$ \tiny{$(3\text{e}-30)$} & $0.160$ \tiny{$(1\text{e}-78)$} & $0.838$ \tiny{$(3\text{e}-107)$} \\
\hline
\end{tabular}
\caption{QbT results for selected experiments. The bottom line has results from the HEM-GMM system. Numbers in brackets are p-values of t-test comparing to the leading representation in the measure, whose score is marked in bold.}
\label{tab:qbt_compare}
\end{table*}

\Cref{tab:qbt_compare} presents the three QbT measures for selected representations, and the generative model alternative (HEM-GMM) as baseline. For each measure, the leading system is marked in bold, and the other systems are compared to it by 2-tailed paired t-test between the two arrays of per-fold-per-tag scores ($N=2905$). The p-values of the t-tests are written in parenthesis.

\subsection{Query-by-example results}

\begin{figure}
    \centering
    \begin{subfigure}[b]{0.5\textwidth}
        \includegraphics[width=\textwidth]{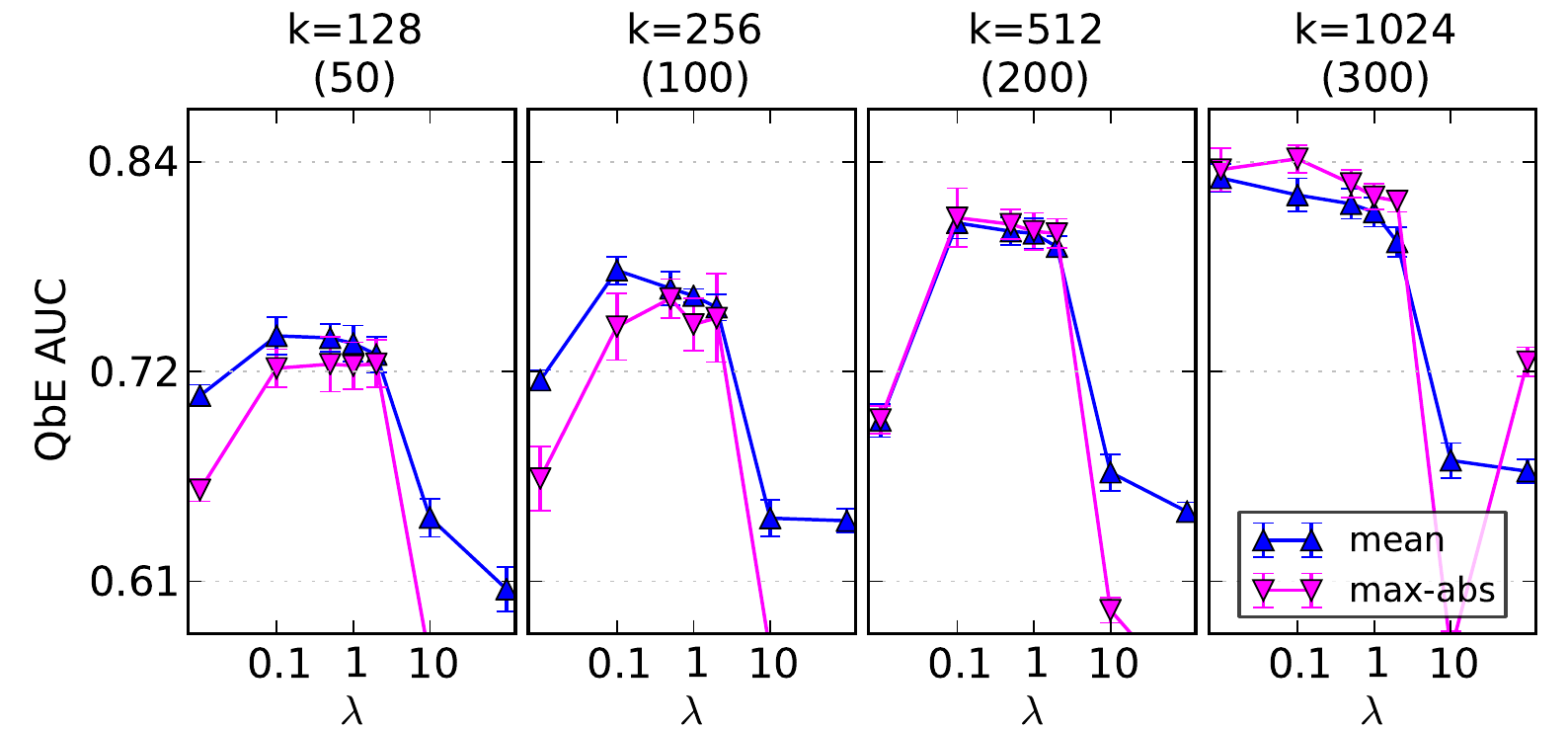}
        \caption{QbE with the LASSO}
\label{fig:query-by-example_sparsity__mel_admm}
    \end{subfigure}
    \begin{subfigure}[b]{0.5\textwidth}
        \includegraphics[width=\textwidth]{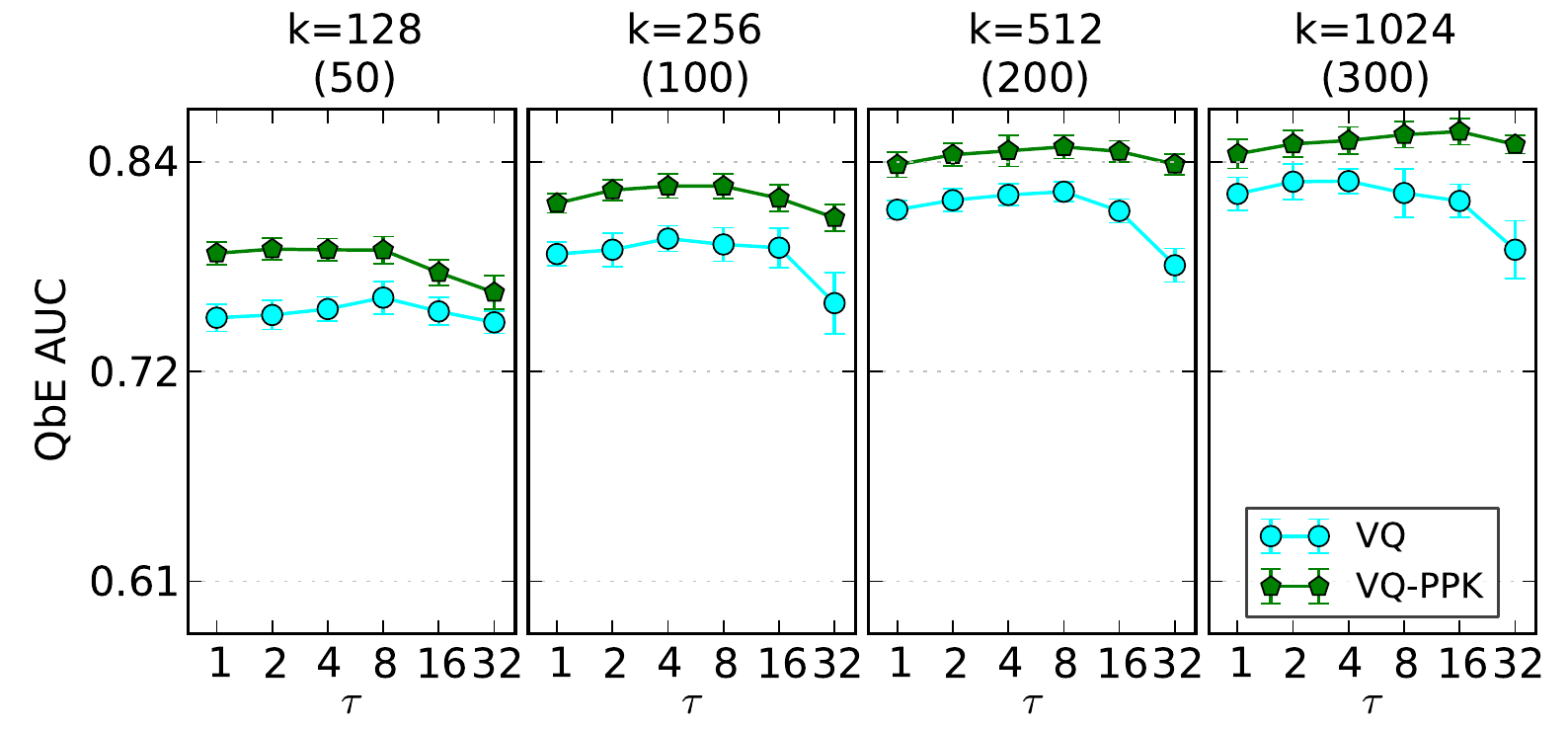}
        \caption{QbE with VQ}
\label{fig:query-by-example_sparsity__mel_vq}
    \end{subfigure}
    \begin{subfigure}[b]{0.5\textwidth}
        \includegraphics[width=\textwidth]{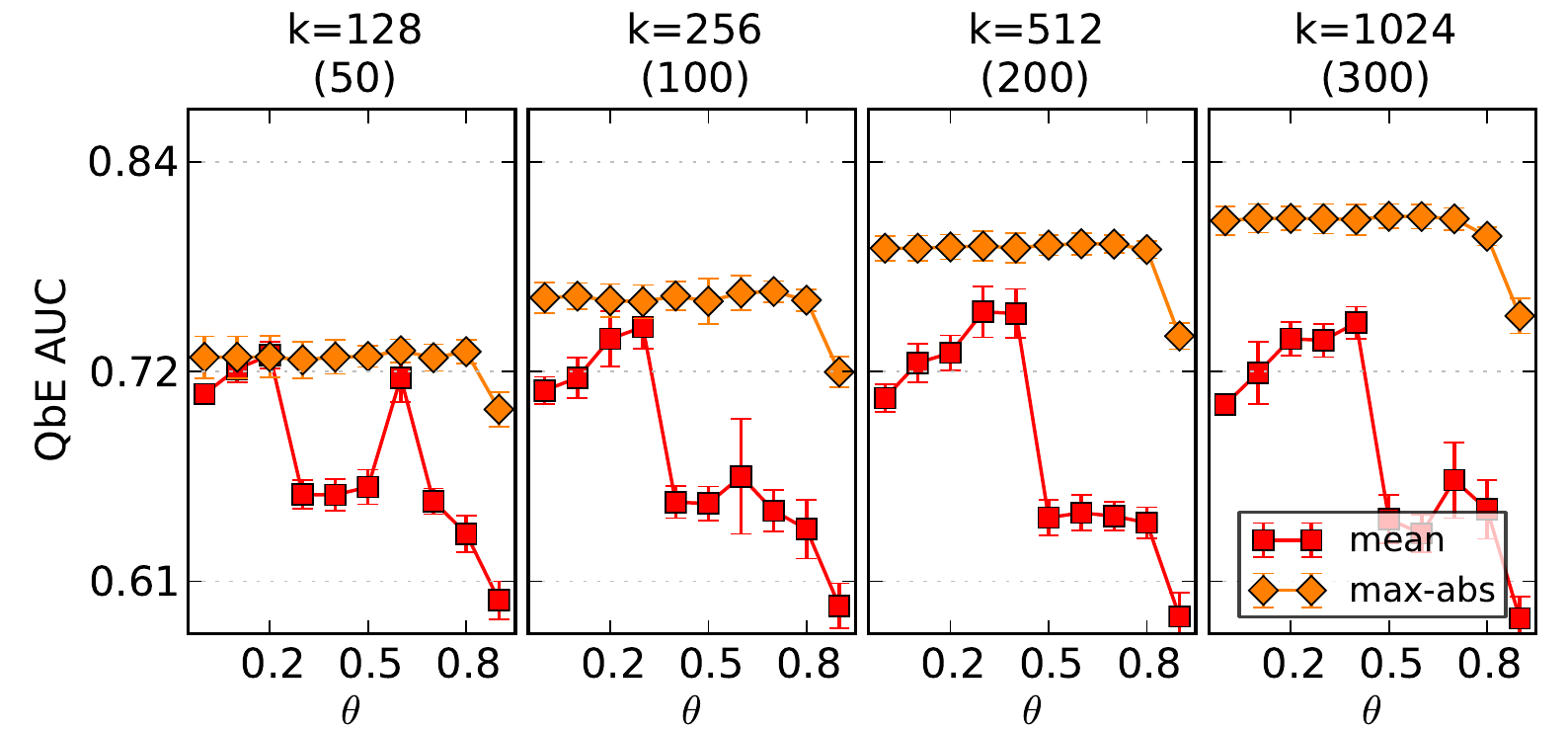}
        \caption{QbE with CS}
\label{fig:query-by-example_sparsity__mel_lf}
    \end{subfigure}

\caption{Query-by-example with different encoders. Effect of pooling or PPK-transformation (shape) and sparsity parameter (x-axis): $\lambda$ (log-scale) for the LASSO (a), $\tau$ (log-scale) for VQ (b) and $\theta$ for CS (c). Error bars indicate one standard deviation over the 10 splits. For each subplot the number beneath the codebook size $k$ is the reduced dimension used for QbE.}
\label{fig:qbe}
\end{figure}

Next, we examine the performance of the query-by-example task (AUC) for the various song representations. \Cref{fig:query-by-example_sparsity__mel_admm,fig:query-by-example_sparsity__mel_vq,fig:query-by-example_sparsity__mel_lf} show the query-by-example results for the three encoding methods. The PCA dimension chosen for each $k$ is written in parenthesis in the title of each subplot. We also experimented with higher PCA dimensions and got similar results (the performance values were slightly higher, but the comparisons among encoders or encoding parameters was the same. See supplementary material).

As expected all encoding methods show improvement with increasing codebook size $k$. For the LASSO (\Cref{fig:query-by-example_sparsity__mel_admm}), again, we see the sensitivity to $\lambda$ (this time mean pooling is also harmed by too low $\lambda$). Unlike for query-by-tag, here there is no strong advantage of max-abs pooling over mean pooling.

For VQ (\Cref{fig:query-by-example_sparsity__mel_vq}) we get partial reproduction of the trends found by McFee \etal~in~\cite{mcfee2012-taslp}: improved performance with increasing codebook size and significant improvement when adding the PPK-transformation. However, since in~\cite{mcfee2012-taslp} the representations were reduced to the estimated effective dimensionality, which was a decreasing function of $\tau$, there was a different effect of $\tau$ than what we find here (where we fix the reduced dimension for a given $k$): where in~\cite{mcfee2012-taslp}, for $k=512,1024$ with PPK increasing $\tau$ seemed to hurt the performance, we show that when PCA is done to a fixed dimension, increasing $\tau$ can maintain a stable performance, and even slightly improve the performance (for both with/without PPK), peaking at around $\tau=8$.

For CS (\Cref{fig:query-by-example_sparsity__mel_lf}), unlike in query-by-tag, there is a significant advantage to max-abs pooling over mean pooling. We again see the damage of over-sparsity: max-abs pooling performance stays stable but decreases after $\theta=0.8$ and mean pooling performance increases with $\theta$ but after peaking early it decreases and stays low.

Both CS and the LASSO are sensitive to the selection of their sparsity parameter: selecting an inappropriate value results in poor performance of the representation. In practical systems such methods require cross validation to select the appropriate parameter value. VQ, on the other hand, is less sensitive to its density parameter $\tau$. This is perhaps due to the fact that $\tau$ directly controls the level of sparsity in the VQ code, whereas for CS and the LASSO the level of sparsity is regularized indirectly. VQ is a stable representation method that can be easily controlled and adjusted with little risk of harming its informative power. VQ consistently achieves highest query-by-example performance (this is also consistent when reducing to a higher PCA dimension. Supplementary material).

\textbf{Comparing both MIR tasks}
\begin{figure}
    \includegraphics[width=0.50\textwidth]{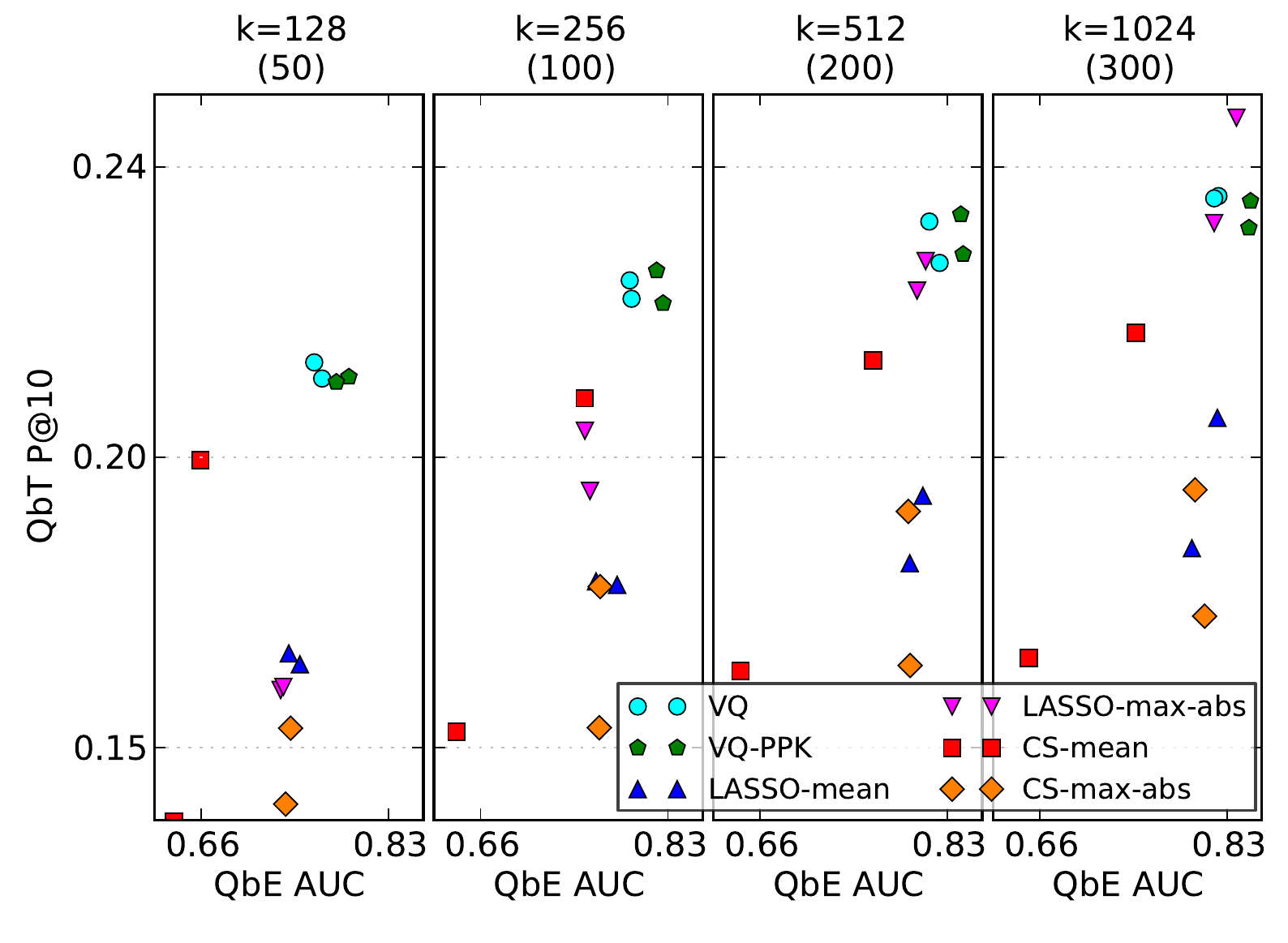}
    \caption{Comparing both MIR tasks: Each point represents a different audio-representation (encoder, parameter, pooling, PPK) and describes its performance in query-by-tag (y-axis) and query-by-example (x-axis). From each encoder-pooling combination the two best performing parameter values are displayed (with same shape). For each subplot the number beneath the codebook size $k$ is the reduced dimension used for QbE.}
\label{fig:comparing_tasks}
\end{figure}

\Cref{fig:comparing_tasks} shows the performance of the same representations in both query-by-tag and query-by-example. The best parameter values from each encoder are presented. The best QbT performance is registered for the LASSO (with max-abs pooling) for $k=1024$, where VQ is slightly behind. However, for QbE VQ consistently leads over the other encoding methods, and the same for QbT with $k<1024$. VQ is a stable and reliable method for both MIR tasks.

\subsection{Encoding runtime}

\begin{figure}
    \includegraphics[width=0.50\textwidth]{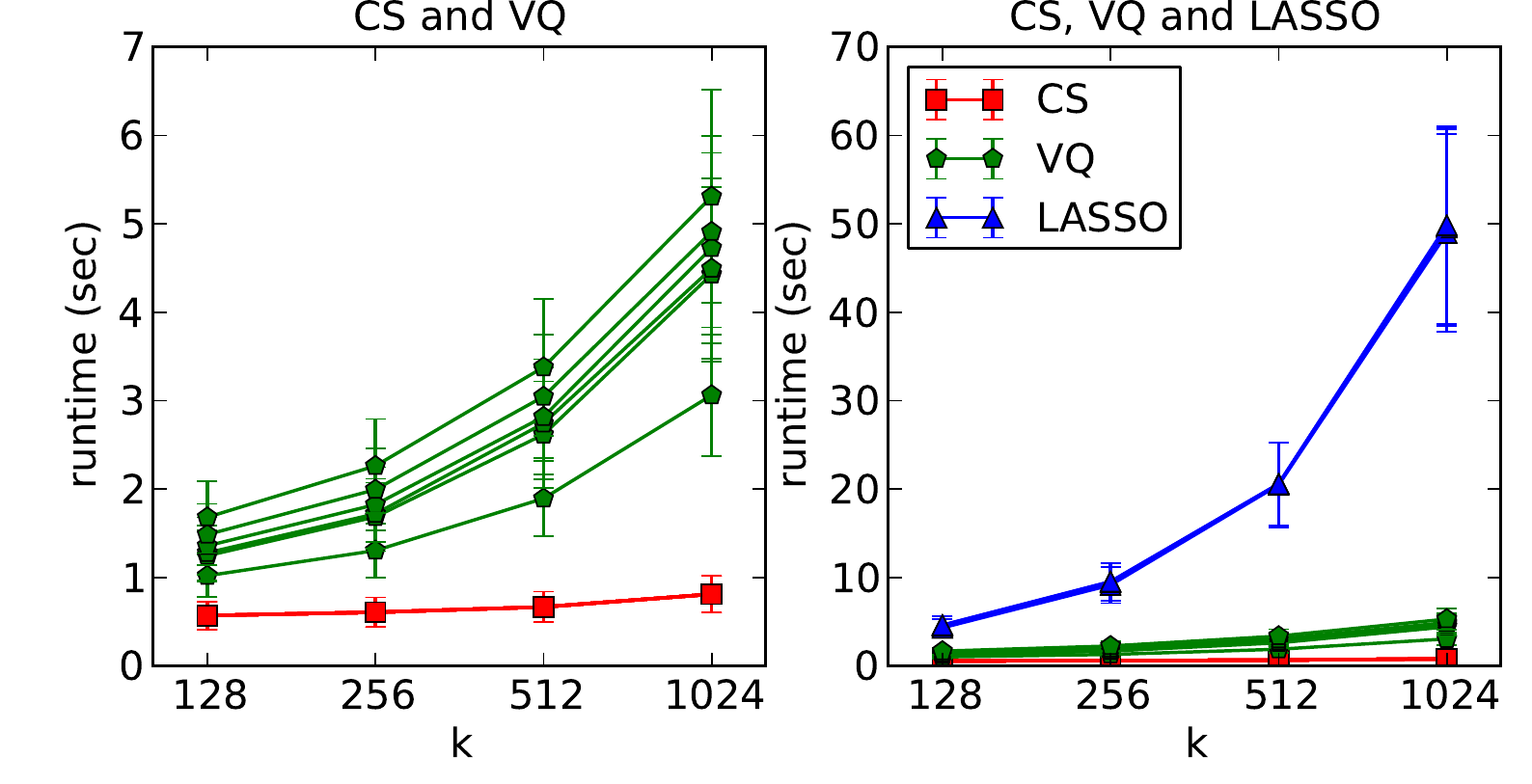}
    \caption{Empirical runtime test. Average runtime for encoding a song as a function of $k$ (log-scale), and standard deviation in error-bars. The left plot is a ``zoom in'' on CS and VQ only. Notice the right plot (containing also the LASSO) has a wider range for y-axis. Multiple points of the same shape represent encodings with different encoding parameter value.}
\label{fig:empirical_runtime}
\end{figure}

As we are searching for practical features and representations for large scale systems, computation resources should also be of consideration when selecting a preferred representation method. We compare the runtime complexity of the three encoding methods, from feature vector $x_t \in \field{R}^d$ to code vector $c_t \in \field{R}^k$:
\begin{itemize}
  \item CS involves multiplying $x_t$ by matrix $D$ ($O(dk)$), computing $\|x_t\|_2$ ($O(d)$) and applying shrinkage to the cosine similarities ($O(k)$), resulting in total complexity of $\bm{T_{\text{CS}}=O(dk)}$.
  \item VQ involves the same matrix-vector multiplication and norm calculation to compute the Euclidean distances. Then $O(c_{\tau,k}k)$ is required to find the $\tau$ closest codewords ($c_{\tau,k}$ is a small number that depends logarithmically on either $\tau$ or $k$, depending on the algorithm used), resulting in total of $\bm{T_{\text{VQ}}=O((d+c_{\tau,k})k)}$.
  \item The ADMM solution for the LASSO is an iterative procedure. Each iterations includes a multiplication of a $(k \times k)$ matrix by a $k$ dimensional vector ($O(k^2)$), a shrinkage function ($O(k)$) and vector additions ($O(k)$), resulting in complexity of $O(k^2)$ per iteration. On top of that, there is $O(dk)$ for once multiplying the dictionary matrix by the feature vector, and there are $M_\epsilon$ iterations, until the procedure converges to $\epsilon$-tolerance, so the complexity for the LASSO encoding becomes $\bm{T_{\text{LASSO}}=O(M_\epsilon k^2 + dk)}$.
\end{itemize}
CS is the lightest encoding method and VQ adds a bit more computation. Recently linear convergence rate was shown for solving the LASSO with ADMM~\cite{hong2012linear}, implying that $M_\epsilon = O(\log{\frac{1}{\epsilon}})$, but even with fast convergence ADMM is still heavier than VQ.\@
This theoretical analysis is verified in empirical runtime measurements, presented in \Cref{fig:empirical_runtime}. We average over the same 50 songs, and use the same computer (PC laptop) with single CPU core. The runtime tests fit a linear dependency on $k$ for CS and for VQ (with slope depending on $\tau$) and a super-linear dependency on $k$ for the LASSO.\@

Using the LASSO (with max-abs pooling) achieves highest performance scores in the query-by-tag task, but the price of runtime requirements is high, and it can be much reduced, by using VQ, while giving up only slightly on query-by-tag performance, and gaining better performance for query-by-example.

\section{Conclusion}\label{sec:conclusions}
We show an advantage to using PCA decorrelation of MFS$\Delta$ features over MFCC.\@ The difference is statistically significant, but small, showing that also the data-agnostic DCT manages to compress music data well.
Increasing the codebook size (up to 1024) results in improved performance for all the encoding methods.
The level of sparsity of the code has an effect (possibly indirect) on performance for all encoding methods, where optimality is achieved with codes that are not too sparse and not too dense. While the LASSO and CS can suffer sharp decrease of performance when adjusting their sparsity parameters, VQ is more robust, having smooth and controlled change in performance when adjusting its density parameter $\tau$.

We find that a simple, efficient encoding method (VQ) can successfully compete with the more sophisticated method (the LASSO), achieving comparable, and even better performance, with much less computing resources. Using \emph{top-$\tau$ VQ with PPK transformation} consistently achieves high performance (almost always beating other methods) in both query-by-tag and query-by-example. It is fast and easy to compute, and it is easily adjustable with its parameter $\tau$. We recommend this representation method as a recipe to be applied to other low-level features, to represent various aspects of musical audio. The resulting representations are concise, easy to work with and powerful for music recommendation in large repositories.

%%%\section*{Acknowledgment}

%%The authors would like to thank...

\ifCLASSOPTIONcaptionsoff
  \newpage
\fi


%\bibliographystyle{IEEEtran}
\bibliographystyle{IEEEtran}
\bibliography{References}

\begin{thebibliography}{10}
\providecommand{\url}[1]{#1}
\csname url@samestyle\endcsname
\providecommand{\newblock}{\relax}
\providecommand{\bibinfo}[2]{#2}
\providecommand{\BIBentrySTDinterwordspacing}{\spaceskip=0pt\relax}
\providecommand{\BIBentryALTinterwordstretchfactor}{4}
\providecommand{\BIBentryALTinterwordspacing}{\spaceskip=\fontdimen2\font plus
\BIBentryALTinterwordstretchfactor\fontdimen3\font minus
  \fontdimen4\font\relax}
\providecommand{\BIBforeignlanguage}[2]{{%
\expandafter\ifx\csname l@#1\endcsname\relax
\typeout{** WARNING: IEEEtran.bst: No hyphenation pattern has been}%
\typeout{** loaded for the language `#1'. Using the pattern for}%
\typeout{** the default language instead.}%
\else
\language=\csname l@#1\endcsname
\fi
#2}}
\providecommand{\BIBdecl}{\relax}
\BIBdecl

\bibitem{tzanetakis2002musical}
G.~Tzanetakis and P.~Cook, ``{Musical genre classification of audio signals},''
  \emph{IEEE Transactions on speech and audio processing}, vol.~10, no.~5, pp.
  293--302, 2002.

\bibitem{Meng:05}
A.~Meng and J.~Shawe-Taylor, ``An investigation of feature models for music
  genre classification using the support vector classifier,'' in \emph{Proc.
  International Society for Music Information Retrieval conference (ISMIR)},
  2005, pp. 604--609.

\bibitem{reed06}
J.~Reed and C.~Lee, ``{A study on music genre classification based on universal
  acoustic models},'' in \emph{Proc. International Society for Music
  Information Retrieval conference (ISMIR)}, 2006, pp. 89--94.

\bibitem{ellis2007classifying}
D.~P. Ellis, ``Classifying music audio with timbral and chroma features,'' in
  \emph{ISMIR 2007: Proceedings of the 8th International Conference on Music
  Information Retrieval: September 23-27, 2007, Vienna, Austria}.\hskip 1em
  plus 0.5em minus 0.4em\relax Austrian Computer Society, 2007, pp. 339--340.

\bibitem{Grosse:2007}
R.~Grosse, R.~Raina, H.~Kwong, and Y.~Ng, A., ``Shift-invariant sparse coding
  for audio classification.''\hskip 1em plus 0.5em minus 0.4em\relax Conference
  on Uncertainty in AI, 2007.

\bibitem{Manzagol:2008}
A.~Manzagol, P., T.~Bertin-Mahieux, and D.~Eck, ``on the use of sparse
  time-relative auditory codes for music.''\hskip 1em plus 0.5em minus
  0.4em\relax International Society for Music Information Retrieval conference
  (ISMIR), 2008.

\bibitem{mandel08}
M.~Mandel and D.~Ellis, ``{Multiple-instance learning for music information
  retrieval},'' in \emph{Proc. International Society for Music Information
  Retrieval conference (ISMIR)}, 2008, pp. 577--582.

\bibitem{Joder:09}
C.~J. S.~Essid and G.~Richard, ``Temporal integration for audio classification
  with application to musical instrument classification,'' \emph{IEEE
  Transactions on Audio, Speech, and Language Processing}, vol.~17, no.~1, pp.
  174--186, 2009.

\bibitem{Hamel:2010}
P.~Hamel and D.~Eck, ``Learning features from music audio with deep belief
  networks.''\hskip 1em plus 0.5em minus 0.4em\relax International Society for
  Music Information Retrieval conference (ISMIR), 2010.

\bibitem{Henaff:2011}
M.~Henaff, K.~Jarrett, K.~Kavukcuoglu, and Y.~LeCun, ``Unsupervised learning of
  sparse features for scalable audio classification,'' in \emph{International
  Society for Music Information Retrieval conference (ISMIR)}, 2011, pp.
  681--686.

\bibitem{Wulfing:12}
J.~Wulfing and M.~Riedmiller, ``Unsupervised learning of local features for
  music classification,'' in \emph{International Society for Music Information
  Retrieval conference (ISMIR)}, 2012, pp. 139--144.

\bibitem{Yeh:12}
C.~M. Yeh, C. and H.~Yang, Y., ``Supervised dictionary learning for music genre
  classification,'' in \emph{ICMR}, 2012.

\bibitem{yeh2013dual}
C.-C.~M. Yeh, L.~Su, and Y.-H. Yang, ``Dual-layer bag-of-frames model for music
  genre classification,'' in \emph{Proc. ICASSP}, 2013.

\bibitem{mandel06}
M.~Mandel, G.~Poliner, and D.~Ellis, ``{Support vector machine active learning
  for music retrieval},'' \emph{Multimedia systems}, vol.~12, no.~1, pp. 3--13,
  2006.

\bibitem{turnbull2008semantic}
D.~Turnbull, L.~Barrington, D.~Torres, and Lanckriet, ``Semantic annotation and
  retrieval of music and sound effects,'' \emph{IEEE Transactions on Audio,
  Speech, and Language Processing}, 2008.

\bibitem{eck08}
D.~Eck, P.~Lamere, T.~Bertin-Mahieux, and S.~Green, ``Automatic generation of
  social tags for music recommendation,'' in \emph{Advances in Neural
  Information Processing Systems}, 2007.

\bibitem{mahieux08}
T.~Bertin-Mahieux, D.~Eck, F.~Maillet, and P.~Lamere, ``Autotagger: a model for
  predicting social tags from acoustic features on large music databases,''
  \emph{Journal of New Music Research}, vol.~37, no.~2, pp. 115--135, June
  2008.

\bibitem{barrington2008combining}
L.~Barrington, M.~Yazdani, D.~Turnbull, and G.~Lanckriet, ``Combining feature
  kernels for semantic music retrieval,'' 2008, pp. 723--728.

\bibitem{tomasik2009}
B.~Tomasik, J.~Kim, M.~Ladlow, M.~Augat, D.~Tingle, R.~Wicentowski, and
  D.~Turnbull, ``{Using regression to combine data sources for semantic music
  discovery},'' in \emph{Proc. International Society for Music Information
  Retrieval conference (ISMIR)}, 2009, pp. 405--410.

\bibitem{coviello2011}
E.~Coviello, A.~Chan, and G.~Lanckriet, ``{Time Series Models for Semantic
  Music Annotation},'' \emph{Audio, Speech, and Language Processing, IEEE
  Transactions on}, vol.~19, no.~5, pp. 1343--1359, July 2011.

\bibitem{Nam:12}
J.~Nam, J.~Herrera, M.~Slaney, and J.~Smith, ``Learning sparse feature
  representations for music annotation and retrieval,'' in \emph{International
  Society for Music Information Retrieval conference (ISMIR)}, 2012, pp.
  565--570.

\bibitem{ellis2013bag}
K.~Ellis, E.~Coviello, A.~Chan, and G.~Lanckriet, ``A bag of systems
  representation for music auto-tagging,'' \emph{IEEE Transactions on Audio,
  Speech, and Language Processing}, 2013.

\bibitem{foote1997content}
J.~T. Foote, ``Content-based retrieval of music and audio,'' in \emph{Voice,
  Video, and Data Communications}.\hskip 1em plus 0.5em minus 0.4em\relax
  International Society for Optics and Photonics, 1997, pp. 138--147.

\bibitem{logan2001music}
B.~Logan and A.~Salomon, ``A music similarity function based on signal
  analysis,'' in \emph{IEEE International Conference on Multimedia and Expo},
  2001, pp. 745--748.

\bibitem{aucouturier02}
J.~Aucouturier and F.~Pachet, ``Music similarity measures: What's the use?'' in
  \emph{Proc. International Society for Music Information Retrieval conference
  (ISMIR)}, 2002, pp. 157--163.

\bibitem{slaney08}
M.~Slaney, K.~Weinberger, and W.~White, ``Learning a metric for music
  similarity,'' in \emph{Proc. International Society for Music Information
  Retrieval conference (ISMIR)}, 2008, pp. 313--318.

\bibitem{hoffman08content}
M.~Hoffman, D.~Blei, and P.~Cook, ``{Content-based musical similarity
  computation using the hierarchical Dirichlet process},'' in \emph{Proc.
  International Society for Music Information Retrieval conference (ISMIR)},
  2008, pp. 349--354.

\bibitem{yoshii2008efficient}
K.~Yoshii, M.~Goto, K.~Komatani, T.~Ogata, and H.~G. Okuno, ``An efficient
  hybrid music recommender system using an incrementally trainable
  probabilistic generative model,'' \emph{Audio, Speech, and Language
  Processing, IEEE Transactions on}, vol.~16, no.~2, pp. 435--447, 2008.

\bibitem{mcfee2012-taslp}
B.~McFee, L.~Barrington, and Lanckriet, ``Learning content similarity for music
  recommendation,'' \emph{IEEE Transactions on Audio, Speech, and Language
  Processing}, vol.~20, no.~8, pp. 2207--2218, October 2012.

\bibitem{logan2000mel}
B.~Logan, ``{Mel frequency cepstral coefficients for music modeling},'' in
  \emph{Proc. International Society for Music Information Retrieval conference
  (ISMIR)}, vol.~28, 2000.

\bibitem{bertin2012large}
T.~Bertin-Mahieux and D.~P. Ellis, ``Large-scale cover song recognition using
  the 2d fourier transform magnitude,'' in \emph{Proceedings of the 13th
  International Conference on Music Information Retrieval (ISMIR 2012)}, 2012.

\bibitem{hamel2011temporal}
P.~Hamel, S.~Lemieux, Y.~Bengio, and D.~Eck, ``Temporal pooling and multiscale
  learning for automatic annotation and ranking of music audio.''\hskip 1em
  plus 0.5em minus 0.4em\relax International Society for Music Information
  Retrieval conference (ISMIR), 2011.

\bibitem{mckinney2003features}
M.~McKinney and J.~Breebaart, ``{Features for audio and music
  classification},'' in \emph{Proc. International Society for Music Information
  Retrieval conference (ISMIR)}, 2003, pp. 151 --158.

\bibitem{flexer2006probabilistic}
A.~Flexer, F.~Gouyon, S.~Dixon, and G.~Widmer, ``{Probabilistic combination of
  features for music classification},'' in \emph{Proc. International Society
  for Music Information Retrieval conference (ISMIR)}, 2006, pp. 111--114.

\bibitem{berenzweig04}
A.~Berenzweig, B.~Logan, P.~W. Ellis, D., and B.~Whitman, ``A large-scale
  evaluation of acoustic and subjective music-similarity measures,''
  \emph{Computer Music Journal}, vol.~28, no.~2, pp. 63--76, 2004.

\bibitem{coviello_vaizman2012}
E.~Coviello, Y.~Vaizman, B.~Chan, A., and G.~Lanckriet, ``{Multivariate
  Autoregressive Mixture Models for Music Autotagging},'' in \emph{13th
  International Society for Music Information Retrieval Conference (ISMIR
  2012)}, 2012.

\bibitem{coviello2012}
E.~Coviello, B.~Chan, A., and G.~Lanckriet, ``{The variational hierarchical EM
  algorithm for clustering hidden Markov models},'' in \emph{Neural Information
  Processing Systems (NIPS 2012)}, 2012.

\bibitem{Jebara:04}
T.~Jebara, R.~Kondor, and A.~Howard, ``Probability product kernels,'' \emph{The
  Journal of Machine Learning Research}, vol.~5, pp. 819--844, 2004.

\bibitem{Lyon:2010}
R.~Lyon, M.~Rehn, S.~Bengio, C.~Walters, T., and G.~Chechik, ``Sound retrieval
  and ranking using sparse auditory representations,'' \emph{Neural
  Computation}, vol.~22, no.~9.

\bibitem{Smith:2006}
C.~Smith, E. and S.~Lewicki, M., ``Efficient auditory coding,'' \emph{Nature},
  vol. 439, pp. 978--982, 2006.

\bibitem{Yang:12}
Y.~Yang and M.~Shah, ``Complex events detection using data-driven concepts,''
  in \emph{ECCV}, 2012, pp. 722--735.

\bibitem{Coates:2011}
A.~Coates and A.~Y. Ng, ``The importance of encoding versus training with
  sparse coding and vector quantization,'' in \emph{International Conference on
  Machine Learning (ICML)}, 2011.

\bibitem{coates2010analysis}
A.~Coates, H.~Lee, and A.~Y. Ng, ``An analysis of single-layer networks in
  unsupervised feature learning,'' \emph{Journal of Machine Learning (JMLR)},
  vol.~15, p. 48109, 2010.

\bibitem{Tibshirani:96}
R.~Tibshirani, ``Regression shrinkage and selection via the lasso,''
  \emph{Journal of the Royal Statistical Society. Series B (Methodological)},
  vol.~58, no.~1, pp. 267--288, 1996.

\bibitem{Boyd:10}
S.~Boyd, N.~Parikh, E.~Chu, B.~Peleato, and J.~Eckstein, ``Distributed
  optimization and statistical learning via the alternating direction method of
  multipliers,'' \emph{Foundations and Trends in Machine Learning}, vol.~3,
  no.~1, pp. 1--122, 2010.

\bibitem{Mairal:10}
J.~Mairal, F.~Bach, J.~Ponce, and G.~Sapiro, ``Online learning for matrix
  factorization and sparse coding,'' \emph{The Journal of Machine Learning
  Research}, vol.~11, pp. 19--60, 2010.

\bibitem{mcfee10_mlr}
B.~McFee and G.~Lanckriet, ``Metric learning to rank,'' in \emph{Proceedings of
  the 27th International Conference on Machine Learning (ICML'10)}, June 2010.

\bibitem{Tingle10}
D.~Tingle, Y.~E. Kim, and D.~Turnbull, ``Exploring automatic music annotation
  with ``acoustically-objectiv'' tags,'' in \emph{Proc. MIR}, New York, NY,
  USA, 2010.

\bibitem{celma:2010}
O.~Celma, ``Music recommendation and discovery in the long tail.''

\bibitem{jaccard}
P.~Jaccard, ``Etude comparative de la distribution florale dans une portion des
  alpes et des jura,'' \emph{Bulletin del la Soci´et´e Vaudoise des Sciences
  Naturelles}, vol.~37, pp. 547--579, 1901.

\bibitem{hong2012linear}
M.~Hong and Z.-Q. Luo, ``On the linear convergence of the alternating direction
  method of multipliers,'' \emph{arXiv preprint arXiv:1208.3922}, 2012.

\end{thebibliography}

%\begin{biography}[{\includegraphics[width=1in,height=1.25in,clip,keepaspectratio]{mshell}}]{Michael Shell}

\begin{IEEEbiography}{Yonatan Vaizman}
Biography text here.
\end{IEEEbiography}
\begin{IEEEbiography}{Brian McFee}
Biography text here.
\end{IEEEbiography}
\begin{IEEEbiography}{Gert Lanckriet}
Biography text here.
\end{IEEEbiography}

%% insert where needed to balance the two columns on the last page with
%%\newpage


%\vfill


\end{document}